\documentclass[aps,prl,showpacs,superscriptaddress,preprint]{revtex4-1}
\usepackage{graphicx, epsfig, amssymb, amsmath, lineno, subfigure, float,lmodern,setspace}
\usepackage{color}

\begin{document}
\pagenumbering{roman}
\title{Describing synchronization and topological excitations in arrays of magnetic spin torque oscillators through the Kuramoto model}

\author{Vegard Flovik}
\email{vflovik@gmail.com}
\affiliation{Department of Physics, NTNU, Norwegian University of Science and
  Technology, N-7491 Trondheim, Norway}

\author{Ferran Maci\`{a}}
\affiliation{Institut de Ci\`encia de Materials de Barcelona (ICMAB-CSIC), Campus UAB, 08193 Bellaterra, Spain}

\author{Erik Wahlstr\"om}
\affiliation{Department of Physics, NTNU, Norwegian University of Science and Technology, N-7491 Trondheim, Norway}

\date{\today}

\begin{abstract}
The collective dynamics in populations of magnetic spin torque oscillators (STO) is an intensely studied topic in modern magnetism. 
Here, we show that arrays of STO coupled via dipolar fields can be modeled using a variant of the Kuramoto model, a well-known mathematical model in non-linear dynamics. By investigating the collective dynamics in arrays of STO we find that the synchronization in such systems is a finite size effect and show that the critical coupling---for a complete synchronized state---scales with the number of oscillators. Using realistic values of the dipolar coupling strength between STO we show that this imposes an upper limit for the maximum number of oscillators that can be synchronized. Further, we show that the lack of long range order is associated with the formation of topological defects in the phase field similar to the two-dimensional XY model of ferromagnetism. Our results shed new light on the synchronization of STO, where controlling the mutual synchronization of several oscillators is considered crucial for applications

\end{abstract}

\maketitle
The emergence of coherent phases of interacting oscillators is at the foundation of the cooperative functioning of a wealth of different systems in nature \cite{biology}. Examples of collective behavior can be chosen within a wide range of systems such as laser arrays \cite{laser}, Josephson junctions \cite{josephson}, chemical reactions \cite{chemical}, synchronously flashing firefly populations \cite{biology}, disease spreading \cite{syncron}, or cortical oscillations in the brain \cite{neuroscience,neuroscience2}. Science has sought mathematical models for understanding collective phenomena in large populations of oscillators that were tractable both analytically and numerically.

The Kuramoto model is a well known mathematical model in non-linear dynamics that describes large systems of coupled phase oscillators \cite{Kuramoto}. The model, with a remarkable simplicity, has been used to describe the essential features of collective excitations in a vast set of biological and physical phenomena \cite{Kuramoto,Kuramoto2,Kuramoto3,Kuramoto4,Kuramoto5,Kuramoto6,Kuramoto7,Kuramoto8,Kuramoto9}. Although the Kuramoto model originally described oscillators interacting all-to-all with the same strength, variations of the model have been used to describe systems with phase offset and time delays in the couplings, other topologies like one-dimensional structures with local couplings etc. (see e.g. ref. \cite{Kuramoto2} for an overview of extensions of the Kuramoto model). 
In particular, two-dimensional Kuramoto networks with diffusive local coupling accept solutions consisting in waves, spirals and many other patterns \cite{topology2}.

Understanding the collective behavior in oscillator networks is also an intensely studied topic in modern magnetism: the synchronization of spin torque oscillators (STO).
STO are strongly non-linear magnetic oscillators that can be implemented into nanoscale devices working at microwave frequencies, and can be frequency and phase locked to external oscillatory signals or other STO \cite{sto,sto2,sto3,sto4,sto5,sto6,sto7,sto8,sto9,sto10,sto11,sto12,sto13,sto14,sto15}.
They are envisaged to be useful for a variety of advanced magnetic nanodevices, as microwave sources and for signal processing in telecommunication technologies (see e.g. ref. \cite{sto_appl, sto_appl2, bio_computing} and references therein).
STO have also been proposed as possible candidates for a full spintronic implementations of neural networks, based on nano-devices emulating both neurons and synapses \cite{bio_computing,  bio_computing3}.
Building artificial neural networks for computation is an emerging field of research within bio-inspired computing  \cite{sto_appl, sto_appl2, bio_computing,bio_computing2,bio_computing3,bio_computing4,bio_computing5,bio_computing6}, where controlling the collective behavior in oscillator networks is crucial.

In both experimental and theoretical studies, most of the work has been performed for limited number of oscillators. Experimentally, the synchronization of STO has proven to be difficult, and the synchronization of only a few oscillators has been demonstrated \cite{sto13, sto15}. Theoretically, the magnetization dynamics of STO is modeled with the Landau-Lifshitz-Gilbert-Slonzewski (LLGS) equation \cite{LLG,LLGS}, but large number of STO lead to challenging computations caused by the non-local dipolar fields. It is important to consider that in these non-linear systems "more is different", and that the collective  behavior can not be derived simply from the behavior of its individual elements. Thus, a theoretical framework capable to capture the essential dynamics would be ideal to explore those systems.

Here, we show that two-dimensional arrays of STO coupled via dipolar fields can be modeled by a variant of the Kuramoto model. We begin with describing two coupled STO with the Thiele equation \cite{Thiele} and show that for small-amplitude oscillations the system can be described as a simple phase oscillator model. 
Next, we model the interactions for the case of a two-dimensional array of oscillators based on the dipolar coupling and obtain a modified Kuramoto model. Finally we compare the results from our model to the micromagnetic solution of the LLGS equation.
 
We find that the synchronization in two-dimensional arrays of dipolar coupled STO is purely a finite size effect and the critical coupling strength for obtaining a globally synchronized state scales with the number of oscillators N as $\lambda_{\text{crit}} \propto \text{log} (N)$. Using realistic values of the dipolar coupling strength between STO we show that this imposes an upper limit for the maximum number of STO that can be synchronized. 
Further, we study the synchronization transition between the initial formation of locally synchronized clusters and the globally synchronized and phase coherent state and correlate it with a transition in the local order of the system. We also observe the emergence of topological defects and the formation of patterns in the phase field similar to the two-dimensional XY-model of magnetism---suggesting a connection between arrays of STO, systems described by a 2d Kuramoto model and the 2d XY model of statistical mechanics.

\section{Results}
\subsection{From the Thiele equation to the Kuramoto model}
\begin{figure}[t]
\centering
\includegraphics[width=80 mm]{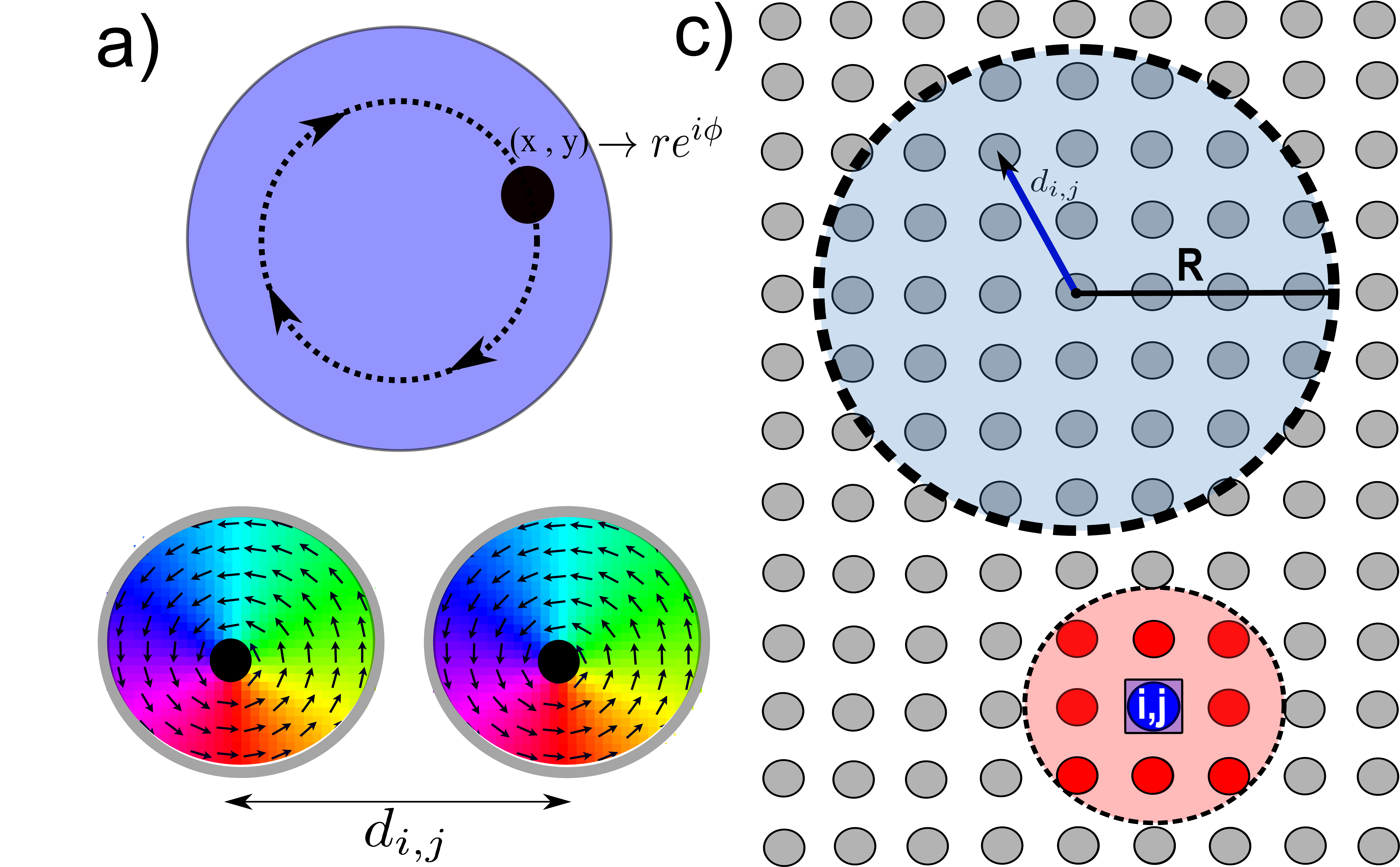}
\caption{\footnotesize a) The gyrotropic motion of the vortex core around the center of the disc can be described by the radius $r$ and phase, $\theta$. b) Two vortex based STO separated by a distance $d_{ij}$, showing the curling magnetization in the disc plane and the location of the vortex core indicated in black. c) Network model for an array of STO. The interaction strength is determined by the spacing $d_{ij}$ and we include interactions within a coupling radius R, indicated by the blue circle in the figure.
The local correlation function $\beta_{i}$ at position $i$ (blue) is given by the degree of synchronization with its neighbors (red).  }
\label{fig:vortices}
\end{figure}

We are considering STO whose free layer ground state configuration is a magnetic vortex. The vortex state is characterized by in-plane curling magnetization, and a small ($\sim 10$ nm) region of the vortex core with out-of-plane magnetization \cite{vortex}. The gyrotropic motion of the vortex core is driven by the injection of a DC spin polarized current through the STO stack, and can be described by a gyration radius $r$ and phase $\theta$, as illustrated in Fig. \ref{fig:vortices}a.
We first model the interaction of two vortices with the Thiele equation with an extra term that accounts for the vortex interaction.
These equations describe the vortices motion given by their coordinates $\bold{X}_{1,2}$ in their self induced gyrotropic mode, and include the spin-transfer-torque (STT) as well as a coupling term \cite{Thiele,Thiele2}.

\begin{equation}\label{eq:thiele}
G(\bold{e_z} \times \dot{ \bold{X}}_{1,2}) - k(\bold{X}_{1,2}) \bold{X}_{1,2} - D_{1,2}  \dot{ \bold{X}}_{1,2} 
 - \bold{F}_{\text{STT} 1,2} -  \bold{F}_{\text{int}}(\bold{X}_{2,1})=0.
\end{equation}

\noindent
Here, $G$ is the gyroconstant,  $k(\bold{X}_{1,2})$ the confining force, $D_{1,2}$ the damping coefficient and $\bold{F}_{\text{STT}}$ the STT.
The interaction between the neighboring STO illustrated in Fig.\ \ref{fig:vortices}b is summarized  by a dipolar coupling term given by $\bold{F}_{\text{int}}=- \mu(d) \bold{X}_{2,1}$, where $\mu(d)$ describes the interaction strength as a function of the separation $d$  between the two STO.

Assuming a small difference in the nominal frequencies of two coupled STO described by Eqs.\ (\ref{eq:thiele}), one can linearize the set of equations following the approach by Belanovsky \textit{et. al.} \cite{sto10}, showing that the dynamics of the phase difference between the STO can be described by Adler`s equation \cite{Adler}. 
Following these approximations, the set of equations reduce to that of two coupled phase oscillators $\theta_1$ and $\theta_2$: (see 'Supplementary information' for details). 

\begin{equation}\label{eq:thiele2}
\dot{\theta_1}=\omega_1 + \lambda \sin (\theta_2 - \theta_1), 
\end{equation}

\begin{equation}\label{eq:thiele3}
\dot{\theta_2}=\omega_2 + \lambda \sin (\theta_1 - \theta_2), 
\end{equation}
\noindent
where $ \omega_{1,2} $ are eigenfrequencies of oscillators $\theta_{1,2}$ respectively, and $\lambda$ describes the interaction strength trying to synchronize them.

To check the validity of the approximations, one can compare the results obtained using a simplified phase oscillator model to a numerical solution of Eqs.\ (\ref{eq:thiele}) as well as a micromagnetic solution of the full system using the LLGS equation. This was done by  Belanovsky \textit{et al}. \cite{sto10}, where they found that for a small difference in nominal frequencies, in their case given by a difference in STO disc diameter $\Delta D/D_0 \leq 5 \%$, the synchronization can be qualitatively described using the simplified model. Assuming the error in state-of-the-art fabrication processes is below this limit, the simplified equations are a valid description of the system.

The functional form of Eqs.\ (\ref{eq:thiele2})-(\ref{eq:thiele3}) is the same as that of the well known Kuramoto model \cite{Kuramoto,Kuramoto2}, which is a generalization for the case of an ensemble of weakly coupled phase oscillators. 
Considering the interaction between several STO, we obtain a Kuramoto model where the single oscillator state is described through the dynamic equation of its phase $\theta_i$ due to the interaction with its surrounding oscillators $\theta_j$:

\begin{equation}\label{eq:kuramoto}
\frac{d \theta_i}{dt} = \omega_i +  \sum_{j \neq i} \lambda_{ij} \sin (\theta_j - \theta_i).
\end{equation}

\noindent
The coupling term is here generalized to include the interaction between several oscillators, determined by the interaction strength $\lambda_{ij}$ between oscillators $\theta_i$  and $\theta_j$. This determines the nature of the interaction, ranging from a global all-to-all coupling where $\lambda_{ij}=\lambda$ for all oscillators, to a local interaction where $\lambda_{ij}=0$ for all but the nearest neighbors.
Here, we are considering the intermediate case of a non-local coupling to mimic the dipolar interaction between neighboring STO.
Starting from a macrodipole approximation for the dipolar energy between two magnetic dipoles $\mu_1$ and $\mu_2$, the average interaction strength is found to decay as $\mu (d) \propto d_{ij}^{-3}$  \cite{dipolar_int}, where $d_{ij}$ is the distance between oscillators $\theta_i$  and $\theta_j$. We thus set the coupling strength to

\begin{equation}\label{coupling}
\lambda_{ij}=
\begin{cases}
  \lambda / d_{ij}^3 & \quad d_{ij} < R \\
 0 & \quad d_{ij} > R, \\
\end{cases}
\end{equation}
\noindent
where we include interactions within a coupling radius $R$, indicated by the blue circle in Fig.\ \ref{fig:vortices}c.
The network model for the STO array is implemented with bi-periodic boundary conditions and
the time evolution of the oscillator phases given by Eq.\ (\ref{eq:kuramoto}) is solved numerically. A small random disorder in the oscillator eigenfrequencies is included by setting $\omega_i = \omega_0 \pm \delta \omega_i$. Here, $\omega_0 = 1$ GHz and $\delta \omega_i$ represents a uniformly distributed random disorder where $\delta \omega_i / \omega_0 \leq 2.5 \%$. The interaction strength is determined by the STO spacing and size, as well as the magnetic material properties \cite{sto14}. Here, the interaction strength has been varied in the range $\lambda=1-20$ MHz, and is in the same range as the interaction strength extracted from micromagnetic simulations for similar STO \cite{sto14}.

In order to evaluate the Kuramoto model as a valid description for arrays of STO, we compare it to a micromagnetic solution of the complete system, accounting for all dynamic dipolar terms (see  'Methods' section).
To compare the Kuramoto model and the micromagnetic solution, we define a suitable order parameter to distinguish disordered and synchronized states.
The phase of the individual oscillators $\theta_i$ is used to define the order parameter $\rho$, describing the phase coherence in a system of $N$ oscillators:

\begin{equation}\label{orderparam}
\rho=\frac{1}{N} | \sum_{j} e^{i \cdot \theta_{j}} |.
\end{equation}

The case $\rho=0$ corresponds to the maximally disordered state, whereas $\rho=1$ represents the state where all oscillators are perfectly synchronized and phase coherent. In addition to the global order parameter $\rho$, we define a local correlation function $\beta$:

\begin{equation}\label{corrfunc}
\beta_{i}=\frac{1}{n} | \sum_{<j>} e^{i \cdot \theta_{j}}|,
\end{equation}
 where the brackets indicate a summation over neighboring oscillators, and $n$ is the number of  neighbors.
$\beta_{i}$ is a measure of the phase correlation of oscillator $\theta_i$ and its neighbors, indicated by the blue and red oscillators in  Fig.\ \ref{fig:vortices}c respectively. If oscillator $\theta_i$ is located within a synchronized cluster, $\beta_{i} \rightarrow 1$. Calculating $\beta$ thus allows for investigating the formation of locally synchronized clusters and the emergence of patterns of synchronized states, which can not be obtained simply from the global order parameter $\rho$.

\subsection{Kuramoto model versus micromagnetic simulations}

\begin{figure*}[t]
\centering
\includegraphics[width=160 mm]{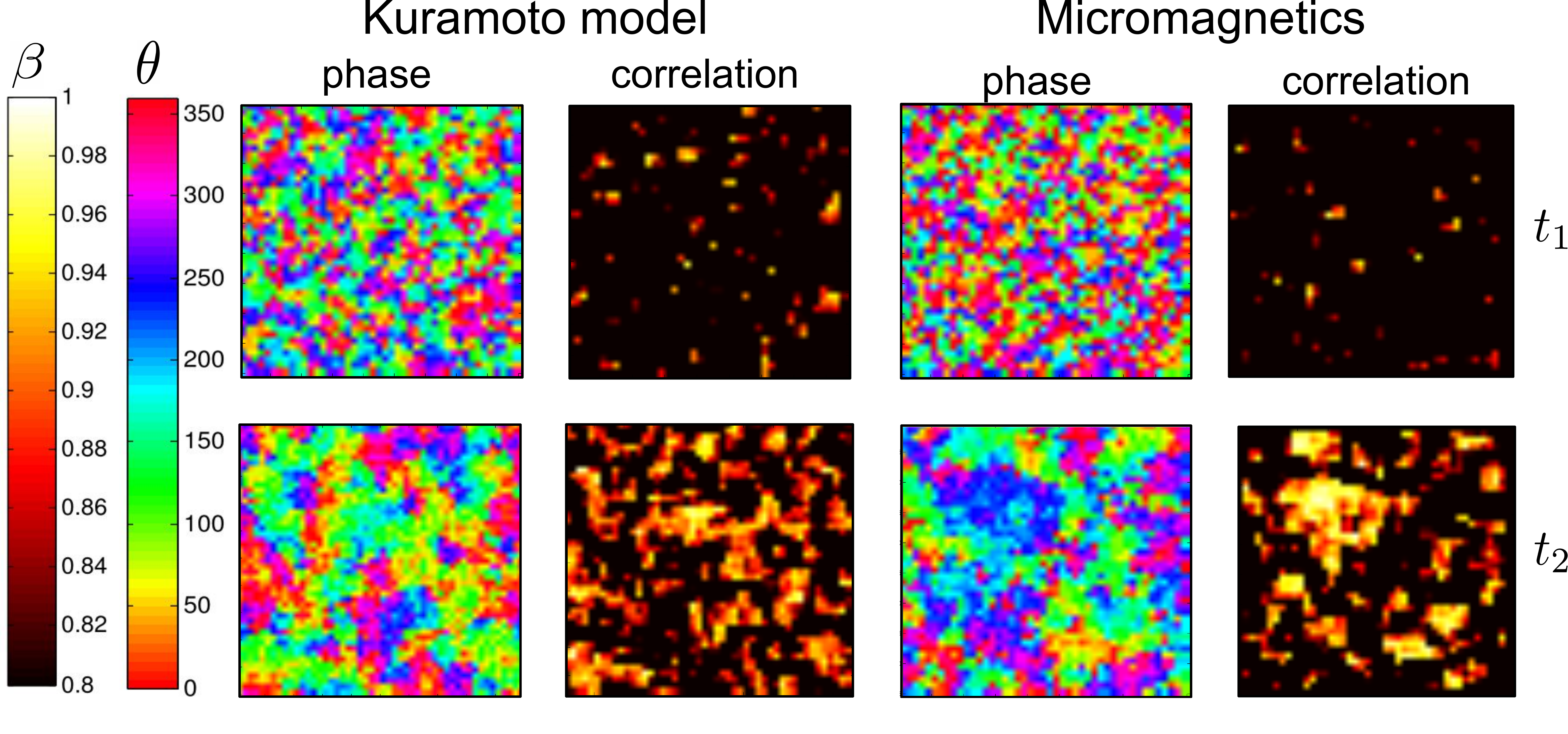}
\caption{\footnotesize Snapshots of the phase map $\theta_{i}$ and local correlation function $\beta_{i}$ for the Kuramoto model and the micromagnetic solution at time $t_1$ and $t_2 > t_1 $ for a network of $45 \times 45$ oscillators.   }
\label{fig:cluster}
\end{figure*}

We now compare the Kuramoto model given by Eq.\ (\ref{eq:kuramoto}) and the full micromagnetic solution of the LLGS equation (see 'Methods' section for details).
Starting from a disordered initial state, we investigate the synchronization dynamics by calculating the time evolution of the phase distribution $\theta_{i}$ and local correlation $\beta_{i}$. As an example, we show in Fig.\ \ref{fig:cluster} snapshots of $\theta_{i}$ and $\beta_{i}$ for the Kuramoto model and the micromagnetic solution for a system of  $45 \times 45$ oscillators at times $t_1$ and $t_2 > t_1$, with random initial phases.
At time $t_1$, one notices the initial formation of small locally synchronized clusters, as seen through both the phase distribution and the bright areas in the correlation maps, where $\beta \rightarrow 1$. As time progress to $t_2$, these clusters grow in size and merge with neighboring clusters.

Comparing the two models, we find that they both show the same behavior. For weak interaction strengths, the system tends to be in a disordered state with no correlation between neighboring oscillators. By increasing the interaction strength above a certain threshold, synchronized clusters begin to form. The oscillators within each cluster are synchronized, but might not be phase coherent with other clusters. This can be seen in the phase maps in Fig.\ \ref{fig:cluster}, where the individual clusters have different phases. As time progresses, the transition from a disordered to a synchronized state is governed by the growth and merging of neighboring clusters, reaching a globally synchronized and phase coherent state for sufficiently strong interactions.

Depending on the interaction strength, the system ends up in either a disordered, partially synchronized or globally synchronized state. Controlling the interaction strength is thus the key parameter to determine the system behavior. The interaction strength needed to obtain synchronization will depend on the differences in the nominal frequencies of the oscillators \cite{sto14}. Another important consideration, is whether the critical interaction strength also depends on the number of oscillators.
Lee \textit{et al.} \cite{Kuramoto_scaling2} have studied the synchronization in a 2d Kuramoto model with a nearest neighbor interaction.
They showed that the transition to a synchronized state depends strongly on system size, and that the critical coupling strength needed to synchronize scales with the number of oscillators $N$ as $\lambda_{\text{crit}} \propto \text{log}(N)$.
This raises the question if such a scaling law can also be observed in our models: observing the same scaling laws in both the Kuramoto model and the micromagnetic solution would strengthen the suggestion of the Kuramoto model as a valid description of arrays of STO.

\begin{figure*}[t]
\centering
\includegraphics[width=160 mm]{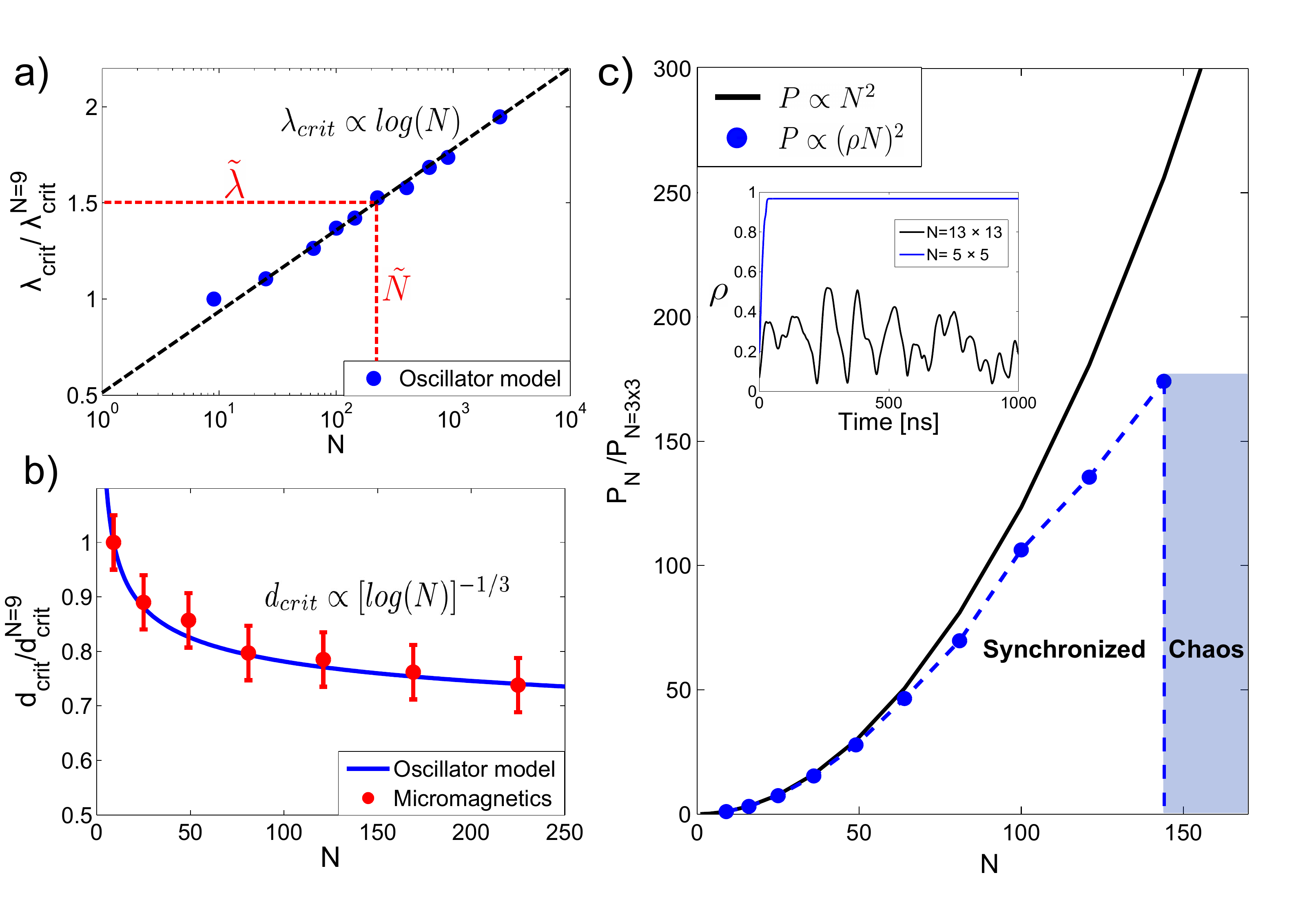}
\caption{\footnotesize a) Critical coupling strength $\lambda_{\text{crit}}$ vs. number of oscillators N in the Kuramoto model, normalized to the case where N $=3 \times 3$. Red dotted line: For a system size $\tilde{N}$, there is a corresponding minimum coupling strength $\tilde{\lambda}$ to obtain a synchronized state, and vice versa.  b) Blue solid line: Expected scaling between critical oscillator spacing, $d_{\text{crit}}$, and number of oscillators when assuming a dipolar interaction decaying as $1/d^3$. Red datapoints: Results from micromagnetic simulations. c) Black solid line: Power output assuming an ideal scaling, $P \propto N^2$. Blue dotted line: Calculated power output for an interaction strength $\tilde{\lambda}= 7.5$ MHz, normalized to the case where N$=3 \times 3$. Inset: Order parameter $\rho$ for $3\times3$ and $13\times 13$ STO respectively, showing the transition from a synchronized state to chaotic behavior as the number of STO is increased. }
\label{fig:scaling}
\end{figure*}

We first consider our Kuramoto model, which has a non-local interaction to mimic the dipolar interaction in arrays of STO. Due to the increased complexity compared to the nearest neighbor model studied by Lee \textit{et al.} \cite{Kuramoto_scaling2}, an analytical derivation of the scaling behavior with system size is to our knowledge still an open question.
To investigate the scaling behavior we thus resort to a numerical solution. We performed simulations with the number of oscillators ranging from $N=9$ to $N=2500$, gradually increasing the interaction strength between each simulation until the system reaches a synchronized state at a critical coupling strength, $\lambda_{\text{crit}}$. 100  simulations were performed for each system size, with different initial oscillator phases and eigenfrequencies. In Fig.\ \ref{fig:scaling}a we show a plot of $\lambda_{\text{crit}}$ vs. number of oscillators, N. The results indicate that the critical coupling strength scales as $\lambda_{\text{crit}} \propto \text{log}(N)$, same as the nearest neighbor Kuramoto model investigated by Lee \textit{et al}. \cite{Kuramoto_scaling2}. 
The main difference compared with the nearest neighbor model is that we include interactions within a coupling radius $R$, as indicated by the blue circle in Fig.\ \ref{fig:vortices}c. The imposed cutoff radius has a physical justification when considering a realistic system, which would inevitably include thermal noise. As the STO we consider are weakly coupled, and the dipolar interaction decay with distance, there will be a limiting spacing where the thermal noise level is comparable to the coupling strength. In our model we observe that the results do not depend qualitatively on the range of the cutoff radius, and that the large scale behavior is dominated by the diffusive coupling. This suggests that insight from studying the analytically tractable nearest neighbor model of Lee \textit{et al.} \cite{Kuramoto_scaling2} might provide valuable insight into the behavior of STO arrays. 

The coupling in the Kuramoto model is defined simply as an interaction strength given by $\lambda$ in Eq.\ (\ref{eq:kuramoto}). In the micromagnetic model the coupling comes from dipolar interactions, determined by the magnetic material properties and the STO spacing.
In order to investigate the scaling behavior in the micromagnetic solution, one thus needs to relate the critical interaction strength to a critical STO spacing.
Following the aforementioned macrodipole approximation that the effective dipolar interaction decay as $1/d^3$, one can relate the coupling strength $\lambda_{\text{crit}}$ to a critical spacing  $d_{\text{crit}}$ between neighboring STO as $d_{\text{crit}} \propto [\text{log}(N)]^{-1/3}$.
Micromagnetic simulations were then performed to obtain $d_{\text{crit}}$ vs. number of oscillators in the range $N=3\times 3$ to $N=15 \times 15$. The results are shown as the red datapoints in
Fig.\ \ref{fig:scaling}b. For comparison, we show as a solid line the expected scaling from the Kuramoto model: $d_{\text{crit}} \propto [\text{log}(N)]^{-1/3}$.
The good agreement between the micromagnetics result and the Kuramoto model indicate that they both follow the same scaling law, strengthening the suggestion of the Kuramoto model as a valid description for arrays of STO.

\subsection{Scaling of output power in arrays of STO}

To investigate the implications of the scaling with system size, we consider a model calculation of the output power as we increase the number of STO. 
For applications of STO as e.g. nanoscale microwave generators, the power output of a single STO is not competitive. Decisive improvement is expected from the synchronization and phase locking of several STO, as this would result in a quadratic increase of the output power, $P \propto N^2$ for $N$ synchronized oscillators.

The output of a single STO can be described by its amplitude and phase, $a_j e^{i \theta_j}$. In our model we have assumed a constant amplitude for all STO and the total amplitude $A$ for an array of STO is given by: $A=\sum_{j} ae^{i \theta_j}$. The power output is proportional to $|A|^2$, and for  N oscillators we obtain $P \propto |aN \frac{1}{N} \sum_{j} e^{i \theta_j}|^2 \propto (\rho N)^2$, from the definition of the order parameter $\rho$ in Eq.\ (\ref{orderparam}). A quadratic scaling in the power output, $P \propto N^2$, implies a perfectly synchronized and phase coherent state, given by $\rho=1$. However, as the coupling strength to obtain a synchronized state scales with the number of STO, this will affect the power output when scaling up to large arrays. 

As an example, we consider a system of STO composed of 200 nm diameter spin valve nanopillars with 15 nm thick Permalloy as the ferromagnetic layer. The average interaction energy can be extracted from micromagnetic simulations, and for an edge-edge spacing of 150 nm the interaction strength is found to be $\tilde{\lambda} \approx 7-8$ MHz \cite{sto14}. To increase the output power, we are now interested in scaling up to a large number of STO. We keep the same STO spacing when scaling up, e.g. keeping $\lambda$ fixed. From Fig. \ref{fig:scaling}a, we see that for an interaction strength $\tilde{\lambda}$, there is a corresponding number of STO, $\tilde{N}$, where $\lambda_{\text{crit}} > \tilde{\lambda}$. This can be illustrated by calculating the total power output as the number of STO is increased, as shown in Fig. \ref{fig:scaling}c. For a small number of STO, we see that the power output follows close to the ideal $N^2$ scaling. However, as the number of STO is increased, the scaling with system size becomes increasingly important. For a certain number of STO, indicated by $\tilde{N}$ in Fig. \ref{fig:scaling}a, the interaction is no longer strong enough to obtain a synchronized state. This is also illustrated in the inset of Fig. \ref{fig:scaling}c, where we plot the order parameter $\rho$ for arrays of $3 \times 3$ and $13 \times 13$ STO respectively, showing the transition from a synchronized state to chaotic behavior as the number of STO is increased above $\tilde{N}$. 

This illustrates the importance of our findings that synchronization in such 2d arrays is purely a finite size effect. The interaction strength is limited by the material properties and STO spacing, and using realistic values of the coupling strength we start to see significant deviations from the ideal $P \propto N^2$ scaling for array sizes larger than $10 \times 10$  STO (see Fig. \ref{fig:scaling}c). This means that in a physical realizable system, the scaling with system size imposes an upper limit for the maximum number of STO that can be synchronized.

\subsection{Topological defects}

That the synchronized and phase coherent state is purely a finite size effect, is similar to that of the classical 2d XY model of magnetism. The Kuramoto model is indeed similar to the 2d XY model \cite{kosterlitz}, where the direction of spin in the  XY model corresponds to the oscillator phase in the Kuramoto model.
In the 2d XY model a long range ordered phase is absent due to the presence of spin wave fluctuations and topological defects. The lack of long range order is a specific case of the Mermin-Wagner theorem in spin systems \cite{Mermin-Wagner}, stating that continuous symmetries cannot be spontaneously broken in systems with sufficiently short range interactions in dimensions $d \leq 2$. The fluctuations preventing long range order in the 2d XY model diverge logarithmically with system size \cite{kosterlitz}, in agreement with the logarithmic scaling observed in our system of STO.

Similar topological defects at the boundaries between locally synchronized clusters have previously been observed in the nearest neighbor Kuramoto model \cite{Kuramoto_scaling2} as well as in other two-dimensional oscillator network models \cite{topology,topology2,topology3,topology4}. In oscillator networks this is associated with the appearance of topological defects in the oscillator phase field, $\theta_{i}$. In the continuum limit this is expressed as:

\begin{equation}\label{topology}
\frac{1}{2 \pi} \oint \nabla \theta(\bold{r},t) \cdot d \bold{l} = \pm n,
\end{equation}
where $d \bold{l}$ is an integration path enclosing the defect, and $n$ is the topological charge.

\begin{figure*}[]
\centering
\includegraphics[width=160 mm]{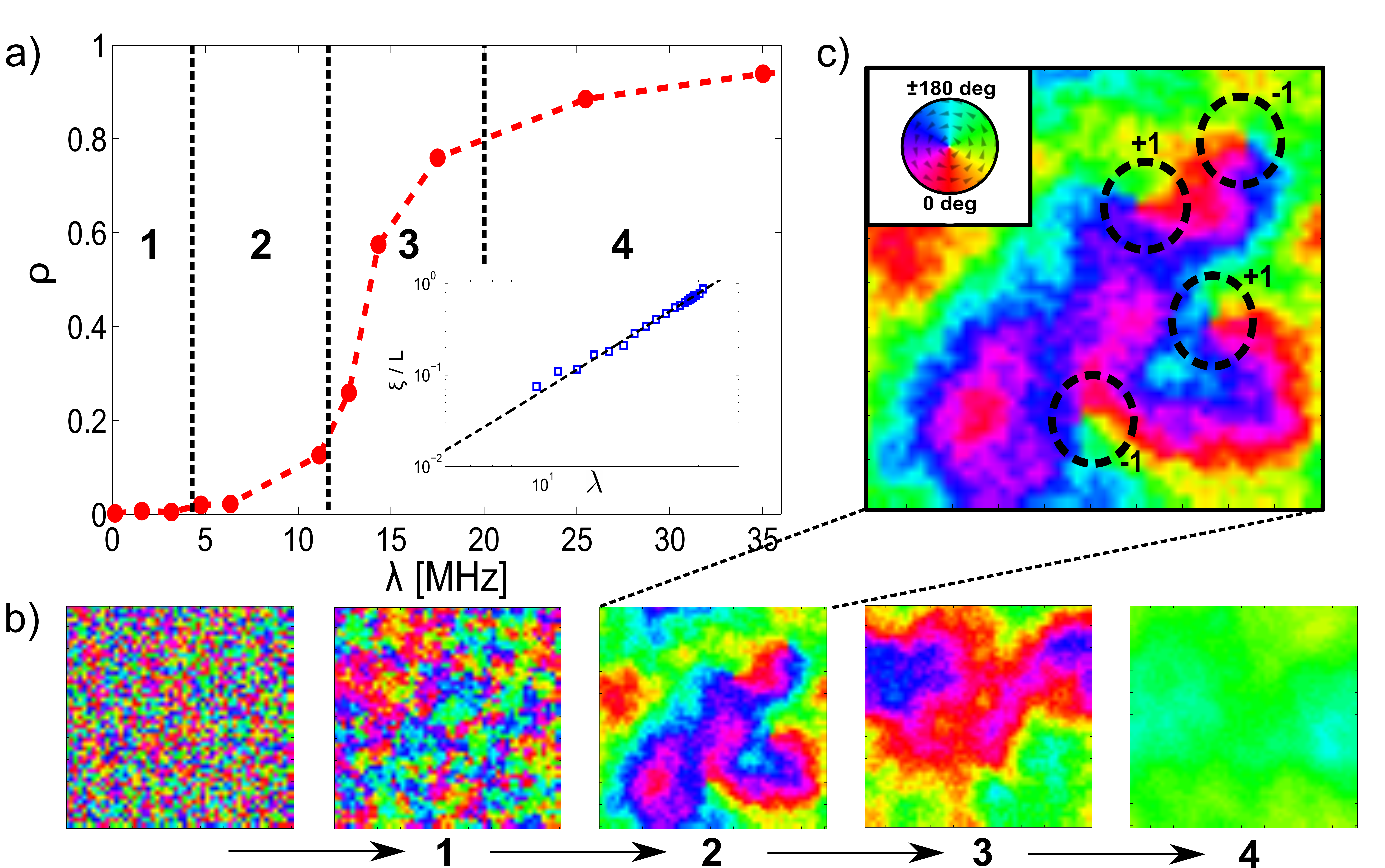}
\caption{\footnotesize  a) Order parameter $\rho$ vs. interaction strength $\lambda$ in the Kuramoto model for a system of $N=50 \times 50$ oscillators, showing the transition from a disordered ($\rho \approx 0$) to a globally synchronized and phase coherent state ($\rho \rightarrow 1$). Inset: Calculated correlation length $\xi$ vs. interaction strength $\lambda$, where the correlation length is normalized to the system size, $\xi/L$ ($N=L \times L$) b) The corresponding phase maps, showing the transition from a disordered state via the formation of locally synchronized clusters (1), vortices (2), spin waves (3) and the globally phase coherent state (4). c) Zoom in of phase map for regime 2, showing the appearance of 4 vortices of charges $\pm 1$ respectively, as defined in Eq.\ (\ref{topology})  }
\label{fig:topology2}
\end{figure*}

Such topological features are observed also in our Kuramoto model for arrays of STO. The presence of vortices in the phase field is more pronounced as the system size increases. As an example we here consider an array of $50 \times 50$ oscillators. The disorder in the system is kept constant (given by the difference in the nominal frequencies of the oscillators) and  the interaction strength $\lambda$ is varied, acting as the inverse temperature: as coupling increases, the system becomes more ordered. Starting from a disordered initial state and varying the interaction strength between each simulation, we observe 4 different regimes:
For weak interaction strengths we observe the formation of locally synchronized clusters, where cluster sizes increase with interaction strength.
Apart from the localized clusters there is no long range order in the system (indicated as regime 1 in Fig.\ \ref{fig:topology2}). Increasing $\lambda$ above a certain threshold, we enter regime 2. Here we observe the formation of vortices in the phase field, and as an example we show in Fig.\ \ref{fig:topology2}c a state with 4 vortices. The topological charge is conserved in the system, and two vortices of charge $\pm 1$ respectively is present.

 In both regime 1 and 2 long range order in the system is absent and $\rho \approx 0$, as indicated in Fig.\ \ref{fig:topology2}a. (that $\rho > 0$ here is a result of fluctuations due to the finite array size). Increasing $\lambda$ further we enter regime 3, where the transition from regime $2 \rightarrow 3 $ is governed by vortex annihilation processes (see 'Supplementary information'). Here there are no topological defects  in the phase field, and the lack of global phase coherence is due to spin waves in the phase field where the oscillator phases change smoothly across the array (regime 3 in Fig.\ \ref{fig:topology2}b).
For sufficiently strong interactions we enter regime 4. Increasing the interaction strength is analogous to increasing the exchange coupling in a Ferromagnetic system, resulting in a more ordered state. The result here is a gradual suppression of the spin waves in the phase field observed in regime 3 as the interaction strength is increased. Regime 4 is thus characterized as the globally phase coherent state where all oscillators are synchronized and phase coherent (regime 4 in Fig.\ \ref{fig:topology2}b).

The growth of the order parameter $\rho$ with increasing coupling strength $\lambda$ in Fig. \ref{fig:topology2}a resembles that of a phase transition. Previous work have shown that the synchronization transition in the globally coupled Kuramoto model can be described as a phase transition, where the nature of the transition can be of first or second-order depending on the frequency distribution and coupling topology \cite{Kuramoto,Kuramoto_phasetrans}. The Kuramoto model with finite range couplings is less studied, as these systems are difficult to analyze and solve analytically. A study of the locally coupled Kuramoto model on a d-dimensional lattice have shown that the synchronization transition depends strongly on the lattice dimensionality, and indicates d=4 as the lower critical dimension for phase synchronization \cite{Kuramoto_critdim}. This is in agreement with the observed scaling with system size in our model, which indicates that the synchronization transition is purely a finite size effect. 

In order to investigate the synchronization transition in our model further, we  calculate the spatial correlation function for the array of oscillators. 
The correlations decay with distance, and asymptotically the correlation function is given  by:
$\langle \theta (\text{r}) \cdot \theta (\text{R}) \rangle \propto e^{- |r - R|/ \xi}/ |r-R|^{\eta}$.
This describes the correlation between oscillators at positions $r$ and $R$ respectively, and the correlation length $\xi$ is obtained by averaging over all positions $r$ and $R$ in the array (an example is shown in 'Supplementary information' Fig. S2).
From the decay of the correlation function, we obtain the correlation length $\xi$ as a function of the interaction strength $\lambda$.
Conventional phase transitions are accompanied by a diverging correlation length close to the transition. Here, we do not observe a diverging $\xi$ going from the disordered to the phase coherent state ($1 \rightarrow 4$ in Fig. \ref{fig:topology2}), and the correlation length remains finite.
As inset in Fig. \ref{fig:topology2}a we show a log-log plot of the correlation length normalized to the system size, $\xi/L$, ($N=L \times L$). The results indicate a power law relating the correlation length and the interaction strength as $\xi \propto \lambda^{\nu}$, where the exponent for this case was found to be $\nu = 2.1 \pm 0.1$.
This means that the correlation length simply scales with the coupling strength, and the transition between regimes 1-4 in Fig. \ref{fig:topology2} correspond to structures of ever increasing length scales. 
The transition to the phase coherent state ($\rho \rightarrow 1$) occurs when the correlation length approaches the system size L, underlining the finite size effects on the synchronization transition and that the system is not undergoing a conventional phase transition. 
Further investigations of finite size effects, the lack of long ranger order in the Kuramoto model and the connection to the 2d XY model will be the subject of future work. 

\section{Discussion}
To summarize, we have shown that the Kuramoto model provides a good description of arrays of  STO. It provides a simple theoretical model to study large populations of coupled STO, which were previously unaccessible due to the long computation time for a full micromagnetic solution. By investigating the collective dynamics in large arrays of STO, we observed a scaling with system size indicating that the synchronization in arrays of dipolar-coupled STO is purely a finite size effect. The critical coupling strength to obtain a globally synchronized state scales with the number of oscillators, as $\lambda_{\text{crit}} \propto \text{log} (N)$, preventing global synchronization for large system sizes. As a consequence of the scaling with system size, we showed that for realistic values of the dipolar coupling strength between STO this imposes an upper limit for the maximum number of oscillators that can be synchronized. 
Further, we showed that the lack of long range order and scaling with system size is associated with the emergence of topological defects and the formation of patterns in the phase field, similar to that of the 2d XY-model of magnetism.

In the present study we considered dipolar-coupled STO, where the short time delay in the coupling between neighboring oscillators compared to the oscillator frequency means that phase delay in the couplings can be neglected.
However, for other coupling mechanisms, time delay can become significant. Interaction mediated by spin waves provide a different mechanism to obtain synchronization of STO \cite{sto15}, where the finite propagation speed of the spin waves results in a phase offset in the couplings. 
Another recent proposal includes the use of non-local electrical couplings, where the coupling phase can be externally tuned through an electrical delay line \cite{electrical_coupling}.

From a dynamical systems point of view, the study of time delay induced modifications to the couplings is of fundamental interest, as well as of practical relevance for modeling of physical, biological and chemical systems. In such systems, time delay is associated with finite propagation velocity of the couplings via e.g latency times of neuronal excitations, reaction times in chemical systems etc. (see e.g. ref. \cite{Kuramoto2} and references therein). The possibility of designing STO arrays with a defined phase offset in the couplings suggests a real world analog to the more general Sakaguchi-Kuramoto model on a 2d lattice, which allows for a phase lag in the couplings \cite{Sakaguchi_Kuramoto}.

Our study suggests, on the one hand, that the use of models from non-linear dynamics can be useful for describing synchronization of magnetic oscillators and, on the other hand, arrays of STO as a a physical realizable model system for the Kuramoto model on a 2d lattice.

\section{Methods}
\subsection{Micromagnetic model} \label{sec:Micromagn}

The micromagnetic model is defined as arrays of discs, where the system is divided into a grid with a mesh size of 5 nm, limited by the exchange length of the ferromagnetic material (here Py). The volumetric quantities such as the magnetization $\bold{M}$ and effective field $\bold{H}_\text{eff}$ are treated at the center of each cell, whereas coupling quantities such as exchange strength are considered at the faces between cells. 
The numerical solution was obtained using the micromagnetic solver Mumax$^3$ \cite{mumax}, which uses a RKF 45 method to solve the Landau-Lifshitz-Gilbert-Slonzewski (LLGS) equation \cite{LLG,LLGS} given by: 

\begin{equation}\label{eq:LLGS_eq}
 \frac{d \bold{M}}{dt} = \underbrace {- \gamma \bold{M} \times \bold{H_\text{eff}} }_\text{Gyration} + \underbrace{ \frac{ \alpha}{M_s} \left[\bold{M} \times  \frac{d \bold{M}}{dt} \right]}_\text{Damping} - \underbrace{ \frac{\chi}{d}J P(\theta)[\bold{M} \times (\bold{M} \times \bold{m}_f]}_\text{Spin-Transfer Torque} . 
\end{equation}
\noindent
Here, $\gamma$ is the gyromagnetic ratio, $\alpha$ the damping parameter and $M_s$ the saturation magnetization. The spin-transfer torque term is given by $\chi =g \mu_b /(2 M_s^2 e)$, the charge current density $J$ and the free layer thickness $d$. $P(\theta)$ is a polarization function assumed to increase with the relative angle $\theta$ between the magnetization of the free layer and the fixed layer and $\bold{m}_f$ is a unit vector in the direction of the magnetization of the free layer. 

In the model, each disc is composed by a magnetic free layer and a fixed polarizer which generates a perpendicular spin polarization $p_z$. The free layer in the STO is 30 nm thick Py with a disc diameter of 150 nm, and the damping parameter $\alpha$ was set to 0.01.
A small disorder in the eigenfrequencies of the individual STO is included through a random distribution in the saturation magnetization in the range $[865,885]\cdot 10^3$ A/m, resulting in a slight variation of STO eigenfrequencies. 
All discs were initialized with a vortex of same polarity and chirality, and the center-center spacing of the discs was varied to change the interaction strength. The polarizing layers are not included in the model as these layers, being uniformly magnetized in z direction have almost no influence on the vortices motion. The vortex gyration was driven by a DC spin current with a polarization $p_z= 0.3$, and current density $J \approx 4.3 \cdot 10^7$ A/$\text{m}^2$. During the simulations, a static magnetic field of 150 mT was applied along the $z$ direction to set the vortex core polarity.

\section*{Acknowledgements}
This work was supported by the Norwegian Research Council (NFR), project number 216700.
V.F acknowledge partial funding obtained from the Norwegian PhD Network on Nanotechnology for Microsystems, which is sponsored by the Research Council of Norway, Division for Science, under contract no. 221860/F40. 
F.M. acknowledges financial support from the Ram\'{o}n y Cajal program  through RYC-2014-16515 and from the MINECO through the Severo Ochoa Program for Centers of Excellence in R\&D (SEV- 2015-0496).

\section*{Author contributions}
V.F initiated the project, developed the model, performed the calculations/simulations and wrote the manuscript. F. M and E. W supervised the project and provided valuable input during the analysis/discussion of results and writing of the manuscript.    

\section{Competing interests}
The authors declare no competing financial interests.

\widetext
\newpage
\begin{center}
\textbf{\large Supplementary information:}
\end{center}

\setcounter{equation}{0}
\setcounter{figure}{0}
\setcounter{table}{0}
\setcounter{page}{1}
\makeatletter
\renewcommand{\theequation}{S\arabic{equation}}
\renewcommand{\thefigure}{S\arabic{figure}}

\subsection{From the Thiele equation to the phase oscillator model}\label{sec:Thiele_Kuramoto}
We here provide some more details on the derivation of the Kuramoto model starting from the coupled Thiele equation:

\begin{equation}\label{eq:thiele_2}
G(\bold{e_z} \times \dot{ \bold{X}}_{1,2}) - k(\bold{X}_{1,2}) \bold{X}_{1,2} - D_{1,2}  \dot{ \bold{X}}_{1,2} 
 - \bold{F}_{\text{STT} 1,2} -  \bold{F}_{\text{int}}(\bold{X}_{2,1})=0.
\end{equation}

\noindent
Here, $G=-2\pi p M_s h /\gamma$ is the gyroconstant, $p$ is the core polarity, $\gamma$ is the gyromagnetic ratio, $M_s$ is the saturation magnetization and $h$ is the thickness of the ferromagnetic layer. The confining force is given by $k(\bold{X}_{1,2}) = \omega_{01,2} G \left(1 + a \frac{   \bold{X}^2_{1,2}  }{R_{1,2}} \right)$ \cite{Thiele3,Thiele4}, where $R_{1,2}$ are the disc radii and the gyrotropic frequency for disc $1,2$ is $\omega_{01,2}=\frac{20}{9}\gamma M_s h /R_{1,2}$. The damping coeficcient $-D_{1,2}= \alpha \eta_{1,2} G$, where $\eta_{1,2}=\frac{1}{2} \text{ln} \left(\frac{R_{1,2}}{2 l_e} \right) + \frac{3}{8}$. Here, $l_e = \sqrt{ \frac{A}{2 \pi M_s}   }$ is the exchange length given by the exchange stiffness $A$ and the saturation magnetization $M_s$. Assuming a uniform perpendicularly magnetized polarizer layer, $\bold{F}_{\text{STT}}=\pi \gamma a_J M_s h (\bold{X}_{1,2} \times \bold{e_z}) = \varkappa  (\bold{X}_{1,2} \times \bold{e_z})$ \cite{Thiele2}, where the spin torque coefficient is given by $a_J = \hbar p_z J / (2 |e| h M_s)$, $\hbar$ is the Planck`s constant, J is the current density and $e$ is the elementary charge.
The interaction between the neighboring vortices in Fig. \ref{fig:vortices}b is summarized by a dipolar coupling term given by $\bold{F}_{\text{int}}=- \mu(d) \bold{X}_{2,1}$, where $\mu(d)$ describes the interaction strength as a function of the separation $d$  between the STO.
 A study of the dipolar interaction between neighboring vortices has been performed by Araujo \textit{et al}. \cite{dipolar_int}. Starting from a macrodipole approximation for the dipolar energy between two magnetic dipoles $\mu_1$ and $\mu_2$, they show that the average interaction energy can be written as $\langle W_{int} \rangle=\mu_{\textit{eff}}C_1 C_2 X_1 X_2$. Here, $C_i$ and $X_i$ are the chirality and gyration radius respectively and $\mu_{\text{eff}}$ is given by:

\begin{equation}\label{eq:dipolar2_2}
\mu_{\text{eff}}=3 \frac{\pi^2 \chi ^2 R^2 h^2}{2d^3},
\end{equation}
\noindent
where $\chi=2/3$, $R$ is the disc radius, $h$ the thickness and $d$ is the inter-disc spacing. In polar coordinates $(X_{1,2} \cos \theta_{1,2}, X_{1,2} \sin \theta_{1,2})$, the coupled equations for two neighboring vortices from Eq.\ (\ref{eq:thiele_2}) can be written as:

\begin{equation}\label{eq:thieleS2}
\frac{ \dot{X}_{1}}{X_{1}} = \alpha \eta_1 \dot{\theta_1}  - \frac{\varkappa}{G} + \frac{\mu X_2}{G X_1} \sin (\theta_1 - \theta_2)
\end{equation}

\begin{equation}\label{eq:thieleS3}
\dot{\theta_1}= - \frac{k(X_1)}{G} - \alpha \eta_1 \frac{\dot{X}_1}{X_1} - \frac{\mu X_2}{G X_1} \cos (\theta_1 - \theta_2)
\end{equation}

\begin{equation}\label{eq:thieleS4}
\frac{ \dot{X}_{2}}{X_{2}} = \alpha \eta_2 \dot{\theta_2}  - \frac{\varkappa}{G} - \frac{\mu X_1}{G X_2} \sin (\theta_1 - \theta_2)
\end{equation}

\begin{equation}\label{eq:thieleS5}
\dot{\theta_2}= - \frac{k(X_2)}{G} - \alpha \eta_2 \frac{\dot{X}_2}{X_2} - \frac{\mu X_1}{G X_2} \cos (\theta_1 - \theta_2)
\end{equation}

\noindent
One can then show that after a few approximations, the set of equations reduce to that of two coupled phase oscillators. We assume the same gyration radius for both vortices, $X_2=X_1$, and that the steady state vortex gyrotropic radius is close to its mean value, $X_0$. This means that Eq. (\ref{eq:thieleS2}) can be set to zero, as $\dot{X}_{1}=0$, and we obtain: 
 
\begin{equation}\label{eq:thieleS6}
\dot{\theta_1}=\frac{\varkappa}{\alpha \eta_1 G} - \frac{\mu}{\alpha \eta_1 G } \sin (\theta_1 - \theta_2)
\end{equation}

Setting $\dot{X}_{1}=0$ and $X_2=X_1$ also in Eq.\ (\ref{eq:thieleS3}): 

\begin{equation}\label{eq:thieleS7}
\dot{\theta_1}=-\frac{k(X_1)}{G} - \frac{\mu}{G } \cos (\theta_1 - \theta_2)
\end{equation}

We then add Eqs.\ (\ref{eq:thieleS6}) and (\ref{eq:thieleS7}) to obtain:

\begin{equation}\label{eq:thieleS8}
\dot{\theta_1}=\frac{\varkappa - \alpha \eta_1 k(X_1)}{2 \alpha \eta_1 G} - \frac{\mu}{2\alpha \eta_1 G }\left[ \sin (\theta_1 - \theta_2) + \alpha \eta_1 \cos (\theta_1 - \theta_2) \right].
\end{equation}
\noindent
Following the same procedure for vortex nr. 2 and assuming low damping, $\alpha \eta << 1$, we obtain the equations for two coupled phase oscillators $\theta_1$ and $\theta_2$:

\begin{equation}\label{eq:thieleS9}
\dot{\theta_1}=\omega_1 + \lambda \sin (\theta_2 - \theta_1),
\end{equation}

\begin{equation}\label{eq:thieleS10}
\dot{\theta_2}=\omega_2 + \lambda \sin (\theta_1 - \theta_2), 
\end{equation}
\noindent
Where $\omega_{1,2} = \frac{\varkappa - \alpha \eta_{1,2} k(X_{1,2})}{2 \alpha \eta_{1,2} G}$ and $\lambda= \frac{\mu}{2\alpha \eta_{1,2} G }$.
The functional form of Eqs.\ (\ref{eq:thieleS9})-(\ref{eq:thieleS10}) is the same as that of the well known Kuramoto model \cite{Kuramoto,Kuramoto2}, which is a generalization for the case of an ensemble of weakly coupled phase oscillators. 
Considering the interaction between several STO, determined by the interaction strength $\lambda_{ij}$ between oscillators $\theta_i$  and $\theta_j$, 
we obtain a Kuramoto model for a population of N interacting oscillators:

\begin{equation}\label{eq:kuramoto2}
\frac{d \theta_i}{dt} = \omega_i +  \sum_{j \neq i} \lambda_{ij} \sin (\theta_j - \theta_i).
\end{equation}

\subsection{Vortex annihilation processes}\label{sec:vortex_annihilation}

Starting from a disordered initial condition, a number of vortices with $n=\pm 1$ is created initially, depending on the array size. Thermal fluctuations of sufficient amplitude could give rise to vortex unbinding, where free vortices proliferate due to thermal fluctuations. As we do not consider thermal effects, such vortex unbinding is not observed this in our model. Since a vortex is topological, it exists until it meets and annihilates with a vortex of opposite polarity, and the transition from disordered to a synchronized state is governed by vortex annihilation processes.

\begin{figure}[h]
\centering
\includegraphics[width=140 mm]{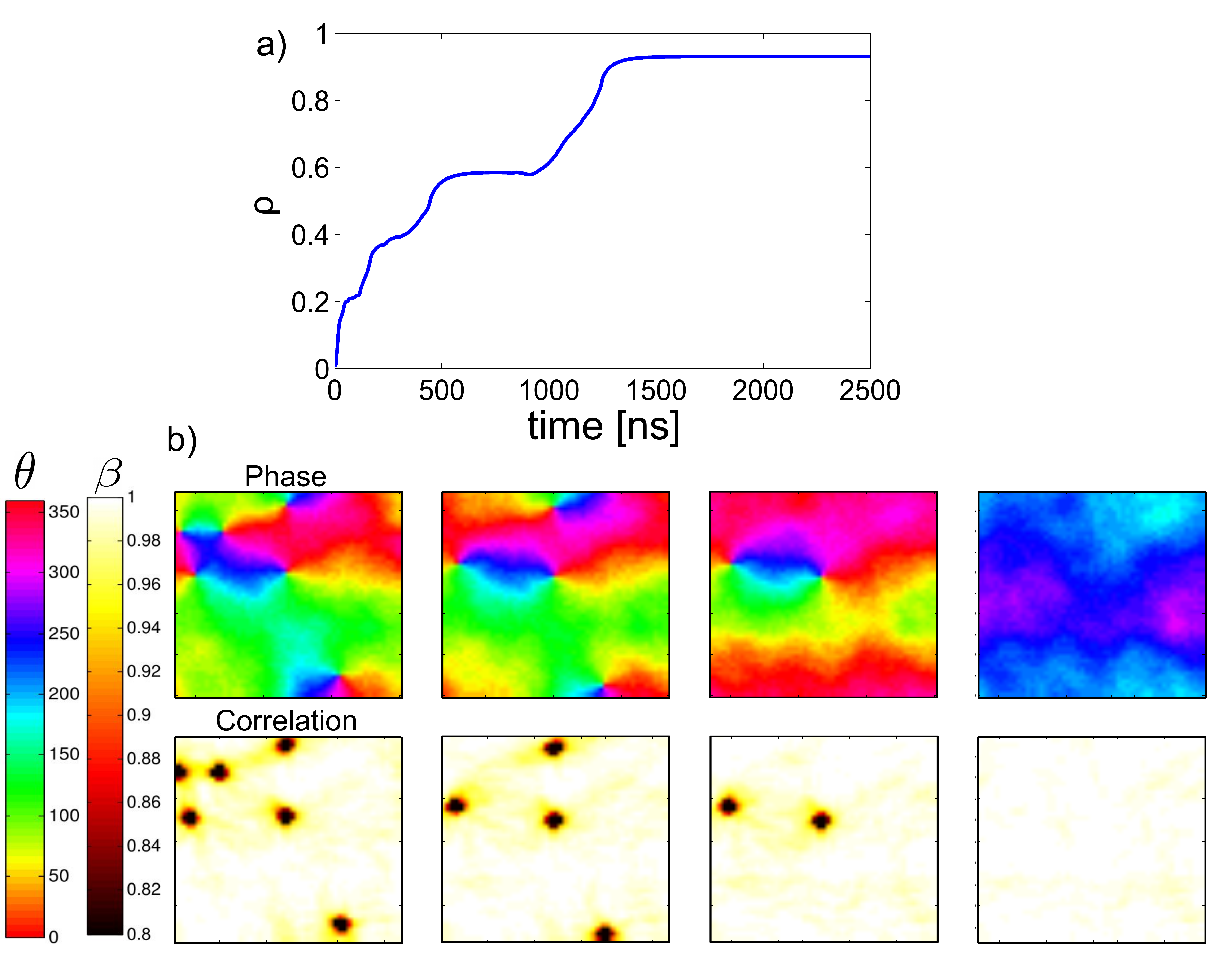}
\caption{\footnotesize a) Order parameter $\rho$ vs. time for an interaction strength of $\lambda=25$ MHz for a system of $50 \times 50$ oscillators, starting from a disordered initial state. b) Snapshots of phase and correlation maps at various timesteps (increasing time from left to right), showing the vortex annihilation processes.   }
\label{fig:topology3}
\end{figure}

In Fig.\ \ref{fig:topology3}a we show the order parameter $\rho$ vs. time, starting from a disordered initial state for a system of $50 \times 50$ oscillators using the Kuramoto model.
The observed jumps in the order parameter correspond to the annihilation of vortices of  charge $\pm 1$. This process is also illustrated in the panels of Fig. \ref{fig:topology3}b, where we show snapshots of the phase map $\theta_{i}$ and local correlation $\beta_{i}$ at various timesteps (with time increasing from left to right). The location and polarity ($n=\pm 1 $) of the vortices can be seen in the phase maps in the upper panels. The position of the vortex core is identified by areas of low correlation ($\beta \rightarrow 0$) between neighboring oscillators, seen as the black spots in the lower panels.
As time progress the vortices annihilate, resulting in a globally synchronized and phase coherent state.

\subsection{Correlation function and correlation length}\label{sec:correlation}

\begin{figure}[h]
\centering
\includegraphics[width=100 mm]{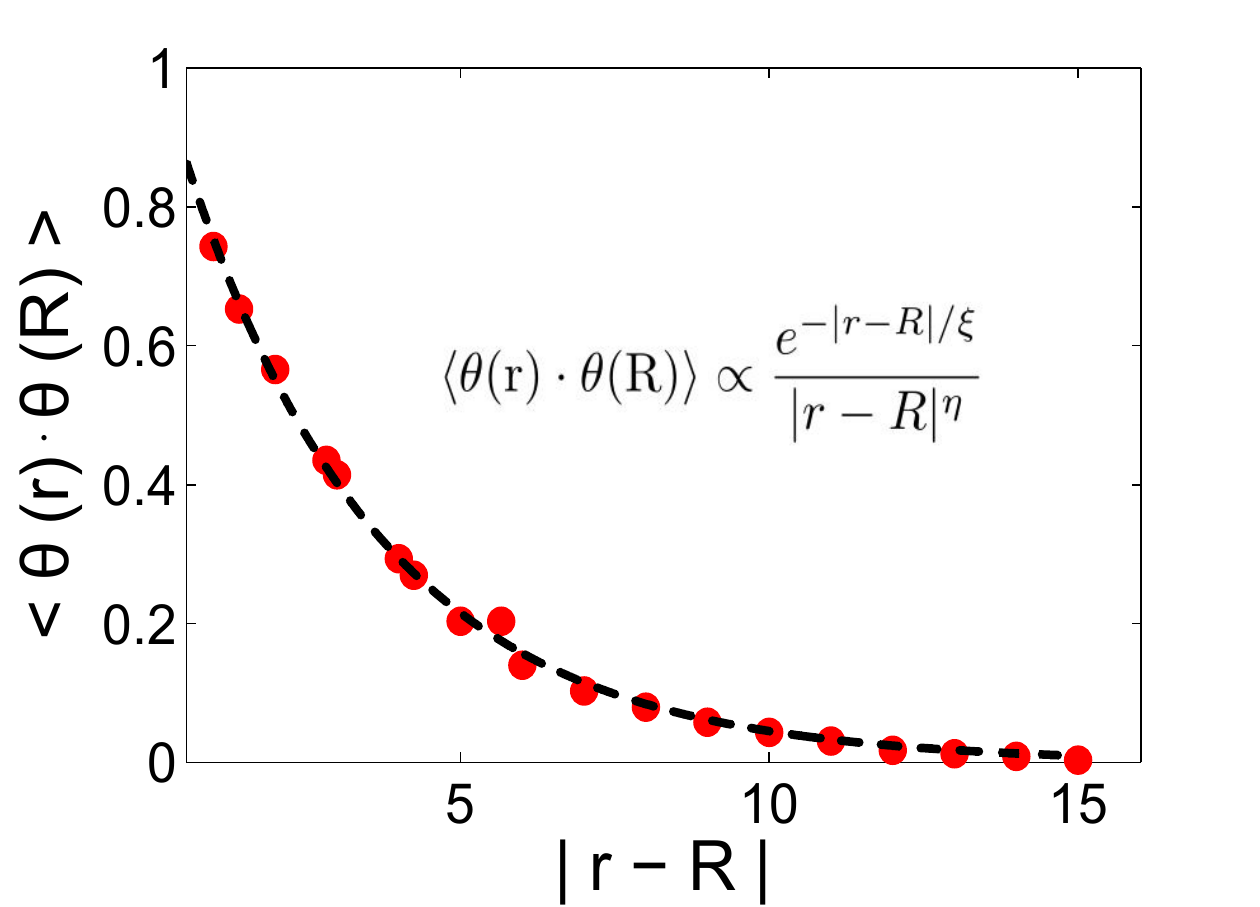}
\caption{\footnotesize Correlation as a function of oscillator spacing, $|r-R|$ for a system of $50 \times 50$ oscillators using the Kuramoto model.}
\label{fig:correlation}
\end{figure}

The spatial correlation function is given asymptotically by:  $\langle \theta (\text{r}) \cdot \theta (\text{R}) \rangle \propto e^{- |r - R|/ \xi}/ |r-R|^{\eta}$. The brackets indicate the correlation between oscillators at positions $r$ and $R$, and the correlation length $\xi$ is obtained by averaging over all positions $r$ and $R$ in the array. An example of the decay of spatial correlations is shown in Fig. \ref{fig:correlation} for a system of $50 \times 50$ oscillators using the Kuramoto model, showing a dominating exponential decay in the correlations for increasing distances between the oscillators. The spacing $|r-R|$ is here expressed in terms of the number of lattice spacings between the oscillators. From the decay of the correlation function, we can then extract the correlation length $\xi$.

\end{document}